\documentstyle[12pt]{article}

\newcommand{\beq}{\begin{equation}}
\newcommand{\eeq}{\end{equation}}
\newcommand{\beqa}{\begin{eqnarray}}
\newcommand{\eeqa}{\end{eqnarray}}
\newcommand{\beqano}{\begin{eqnarray*}}
\newcommand{\eeqano}{\end{eqnarray*}}
\newcommand{\eref}[1]{Eq. (\ref{#1})}
\newcommand{\vo}[1]{{\bf #1}}
\newcommand{\vt}[2]{({\bf #1}+{\bf #2})}
\newcommand{\vmo}[1]{\vert{\bf #1}\vert}
\newcommand{\vmt}[2]{\vert{\bf #1}+{\bf #2}\vert}
\newcommand{\cd}{{\cdot}}
\newcommand{\hq}{\hat Q}
\newcommand{\vhq}{\hat \vo Q}

\begin{document}

\begin{titlepage}

\begin{centering}
\null
\vspace{1cm}
{\bf NON-DEBYE SCREENING IN HIGH TEMPERATURE QCD?}

\vspace{2cm}
{\bf S. Peign\'e \ and \ S.M.H. Wong}

\vspace{1cm}
\em \footnote{Laboratoire associ\'e au Centre National de la
Recherche Scientifique}LPTHE, Universit\'e Paris-Sud, B\^atiment 211,
F-91405 Orsay, France

\end{centering}

\vspace{3cm}

\begin{center}
{\bf Abstract}
\end{center}

\vspace{0.5cm}

We study the non-abelian, gauge invariant, screening potential
between two charges in a quark-gluon plasma. Using Braaten
and Pisarski resummed perturbation theory in temporal axial
gauge, we find a repulsive power screened potential
which behaves as $\sim 1/r^6$ at large distance.
This questions either the physical meaning of the Debye
mass or the known pragmatic treatment of the temporal
axial gauge propagator.

\vspace{2.0cm}
\noindent LPTHE-Orsay 94/46

\noindent May 1994 (revised)

\addtocounter{page}{1}
\thispagestyle{empty}
\pagebreak
\end{titlepage}

\section{Introduction}

Charges in a medium are understood to be screened. At large
distance, this screening is usually taken to be an
exponential \cite{kap} with the screening length given
by the inverse of the Debye mass $m$. In QED, according to linear
response theory, this screening is related to the two-point function
of the photon. This is due to the field strength being
linear in the vector potential. The small
momentum behaviour of this function gives rise to imaginary poles
which leads to the exponentially screened potential, from
which the Debye mass derives its physical significance of
inverse screening length.

In the last few years, the plasmon puzzle or the gauge dependence
of the gluon damping constant has made it clear that it is
necessary to employ resummed perturbation theory of Braaten
and Pisarski \cite{bra&pis} whenever momenta much less than
the temperature T of the system under study are involved.
Since the long distance behaviour of the screening potential
and hence the two-point function is required, it is necessary
to use this resummed theory.

Unlike QED, in QCD only the leading term of the static
time-time component of the gluon self energy, i.e. the Debye
mass squared, calculated in the framework of this
resummed theory, is gauge independent \footnote{In the bare theory
\cite{kaj&kap}, in addition to the gauge invariant leading term,
there is a linear term in the modulus of the external momentum
which is also gauge independent.}. Therefore it is not clear that the
Debye mass is a physical quantity in the non-abelian theory.
By working out the leading correction to the Debye mass in
the resummed theory, Rebhan \cite{reb1,reb2,reb3} argued that in
covariant gauge, it was possible to have a gauge parameter
independent and hence gauge independent Debye mass by defining
\beq m^2 = \lim_{{\vo q}^2 \rightarrow -m^2} \Pi_{00}(0,\vo q) \; ,
\label{massdef}
\eeq
instead of
\beq m^2 = \lim_{\vo q \rightarrow 0} \Pi_{00}(0,\vo q) \; .
\eeq
That is one obtains $m$ self-consistently from the pole of the
propagator.

The leading correction to this gauge independent Debye mass was found
to be
\beq \delta m^2 \sim g m^2 \ln (1/g) \; .
\eeq
The $\ln (1/g)$ is due to the introduction of the magnetic mass as a
infrared cut off in the transverse propagator. As a result,
the correction is rather large. The new
definition \eref{massdef} is possible because the gauge parameter
dependent term in the leading correction in covariant gauge
is proportional to ${\vo q}^2+m^2$. However, it is not clear
why $m$ found this way in a gauge such as covariant gauge
can be readily labelled as the Debye mass: the latter
is generally understood to be the inverse screening length
in a Yukawa type screening potential between charges
and in covariant gauge, the potential is related to more
than just the longitudinal propagator. On the other hand, the
temporal axial gauge (TAG) is the only gauge in which
we have this simplification. Baier and Kalashnikov
\cite{baier&kal} have recently shown, by some unique construction,
that the gauge-invariant definition of the Debye mass might not
extend to TAG: instead of being exponentially screened,
the potential is repulsive with power screening
$\sim 1/r^6$ in TAG \footnote{Another example of non-Debye screening
with a $1/r^{10}$ behaviour in the charge-charge correlation has
been found in \cite{cor&mar}}. It then seems appropriate to
investigate this problem in TAG in the framework of Braaten and
Pisarski resummed theory.

The choice of TAG at finite temperature may be
controversial, because it is known to have problems with
the double pole $1/p^2_0$ when $p_0$ is zero. Nevertheless,
TAG has the advantage that one can relate unambiguously the
potential between two charges to the two-point function.
Also it has been shown in non-resummed theory \cite{hei&kaj&toi}
that the same result for $\Pi_{00}(0,\vo q\rightarrow 0)$ could
be obtained either in Coulomb gauge or in TAG. Basing on this,
we will adopt here a pragmatic attitude and use the ansatz
\eref{ansatz2}. It will be further explained in section
\ref{leadingmom}. We use imaginary time formalism throughout.

\section{Linear Response Theory}

\label{linresp}

The potential of a charge $Q_1$ at ${\vo x}_1$ in the
presence of another charge $Q_2$ at ${\vo x}_2$ is given by
\cite{kaj&kap,kap&toi}
\beq V(\vo r) = {1\over 2}\sum_a \int d^3 \vo x \, \left (
                {\cal E}_1^{a \;\rm eff} (\vo x)
                \cdot {\cal E}_2^a (\vo x) +
                {\cal E}_2^{a \;\rm eff} (\vo x)
                \cdot {\cal E}_1^a (\vo x) \right )
\label{potenteqEE}
\eeq
where the effective field ${\cal E}_1^{a \;\rm eff}$ created by
the non-abelian charge $Q_1^a$ is the sum of the applied field
${\cal E}_1^a$ and the induced field
$\delta \langle\vo E_1^a \rangle$:
\beq {\cal E}_1^{a \;\rm eff}(\vo x) =
     {\cal E}_1^a(\vo x) + \delta \langle \vo E_1^a \rangle
\eeq
${\cal E}_1^a$ is solution of the Gauss' law:
\beq \nabla \cdot {\cal E}^a -g f^{abc}
     {\cal E}^b \cdot {\cal A}^c = Q_1^a
     \delta^3(\vo x -\vo x_1)  \; ,
\label{gauss}
\eeq
where ${\cal A}$ is the vector potential associated with ${\cal E}$.

\eref{potenteqEE} is manifestly gauge invariant since a trace is
being taken over tensor fields (or equivalently colour is being
summed over) and is therefore a physical potential.

When we put the charge $Q_1^a$ in the plasma, the electric field
$\vo E^a_1(\vo x,t)$ in the medium is
coupled to the applied field ${\cal E}_1^a$ through the
coupling Hamiltonian:
\beq H^{\rm ext} (t)= \int d^3 \vo x \,\vo E^a_1(\vo x, t)
                      \cdot {\cal E}_1^a(\vo x) \; .
\eeq
{}From linear response theory, the induced field is
\beqa \delta \langle E^a_i(\vo x, t) \rangle &=& i\int^t_{-\infty}
      dt'\langle [H^{ext}(t'), E^a_i(\vo x, t)] \rangle  \nonumber \\
      &=& i\int^t_{-\infty} dt' \int d^3{\vo x'}
      {\vo {\cal E}}_j^b(\vo x')
      \langle [E^b_j(\vo x', t'), E^a_i(\vo x, t)] \rangle \; .
\label{inducedfield}
\eeqa
In TAG,
\beq E_i^a(\vo x, t) =-\partial_0 A^a_i(\vo x, t) \; ,
\label{EinTAG}
\eeq
the correlator of electric fields $\langle [E,E] \rangle$ can be
easily related to the retarded gluon propagator and the
Gauss' law satisfied by the applied field in the presence of a
static charge $Q_1^a$, is simply \footnote{This is the form of
Gauss' law in any gauge only at leading order.}
\beq \nabla \cdot {\cal E}_1^a = Q^a_1 \delta^3 (\vo x -\vo x_1)
     \; ,
\eeq
where the non-abelian term of \eref{gauss} vanishes for a static
${\vo {\cal E}}$ field. The solution is
\beq {\vo {\cal E}}_i^a(\vo x) = -i Q^a_1 \int
     {{d^3\vo p} \over {(2\pi)}^3} e^{i\vo p\cdot (\vo x-\vo x_1)}
     {p_i \over {\vo p}^2} \; .
\label{gausssolut}
\eeq

In gauges other than TAG, Gauss' law itself is gauge
parameter dependent in a parametrized gauge. The non-abelian
part of \eref{gauss} does not vanish, and thus the solution
${\cal E}^a(\vo x)$ is not of the same simple
form as above and is not necessarily gauge parameter
independent. In parametrized Coulomb gauge, although Gauss' law is not
gauge parameter dependent due to $A^0$ independence of the
gauge fixing term, the solution is not of the simple form of
\eref{gausssolut} because the non-abelian term does not drop
out. Therefore $V(r)$ is not simply related only to
$\langle [E_L,E_L]\rangle$ as in TAG. So that the gauge parameter
dependence of $\langle [E_L,E_L]\rangle$ shown in
\cite{krae&kreu&reb&sch,reb4} does not mean that $V(\vo r)$
is gauge dependent. On the contrary, from the
definition in \eref{potenteqEE}, $V(\vo r)$ is gauge independent.

In TAG, as a consequence of \eref{inducedfield}, (\ref{EinTAG}),
(\ref{gausssolut}), $V(\vo r)$ is directly related to the longitudinal
propagator so we have
\beq V(\vo r) = Q_1\cd Q_2 \int {{d^3 \vo q} \over {(2\pi)^3}}
     {{exp(i\vo q \vo r)} \over {{\vo q}^2 + \Pi_{00}(0,\vo q)}}
\label{QQpotprop}
\eeq
with $\vo r = {\vo x}_1 - {\vo x}_2$. We stress that in other
gauges, this advantage of TAG is gone and $V(\vo r)$ is related
to more than just the two-point function
\footnote{It has been argued \cite{weld} that in the
presence of only one static colour charge, the response
of the vector potential $\langle A\rangle$ must point only in that
direction in colour space, so that a non-abelian term like
$\langle A\rangle \wedge \langle A\rangle$ vanishes. For
this reason, the response field $\langle E\rangle$ and
Gauss' law both become abelian. Using this line
of argument Rebhan \cite{reb2,reb3} argued that the potential
of a static colour charge is given by a form similar
to \eref{QQpotprop}. This is actually flawed because the
response field $\langle E\rangle$ cannot be expressed
directly in terms of the response vector field $\langle A\rangle$.
The response field is actually
$\langle E\rangle = \langle\partial A\rangle-\langle\partial A\rangle
+\langle A \wedge A\rangle$, i.e.
the non-abelian term should be the average of the
cross product and not the cross product of the average, so
it does not vanish. Likewise, the non-abelian term in
Gauss' law cannot be zero.}.

After going through the usual steps, we get for the colour singlet
quark-antiquark potential \cite{kaj&kap},
\beq V(r) = -{{N^2-1} \over {2N}} {{g^2(T)} \over {2\pi^2 r}}
     \int^\infty_0 {{q dq \sin qr} \over {q^2 + \Pi_{00}(0,q)}} \; .
\label{potent}
\eeq
{}From now on, $\Pi_{00}$ is the temperature dependent part of $\Pi_{00}$.

If $\Pi_{00}(0,q)$ is not even in $q$ as found in
\cite{baier&kal,gale&kap}, we have to evaluate
the integral by closing the contour in the first quadrant of the
complex $q$ plane. Contributions from possible poles or branch cuts
in the first quadrant away from the real axis are exponentially
suppressed as $r \rightarrow \infty$.
\eref{potent} can be rewritten as
\beq V(r \gg m^{-1}) \approx {{N^2-1} \over {2N}}
     {{g^2(T)} \over {2\pi^2 r}} {\rm Im} \int^\infty_0
     {{z dz e^{-rz}} \over {-z^2 + \Pi_{00}(0,q= iz)}} \; .
\label{Eucpotent}
\eeq
\eref{Eucpotent} shows that we only need the small $q$ limit of
$\Pi_{00}(0,q)$.

\section{Braaten and Pisarski Resummation in Temporal Axial Gauge}

In TAG, the gluon propagator is
\beqa D_{00} (p) & = & 0 \; ,\mbox{\hskip 2.0cm}
      D_{0i} (p) \; = \; 0 \nonumber \\
      D_{ij} (p) & = & -{1 \over {p^2-G(p)}}
             \left ( \delta_{ij}-{{p_i p_j} \over {\vo p}^2} \right )
                       -{1 \over {p^2-F(p)}}
                   {p^2 \over {p_0}^2} {{p_i p_j} \over {\vo p}^2}
\label{axiprop}
\eeqa
from which the effective or resummed propagator $^*\!D_{\mu \nu}(p)$
can be constructed by replacing G(p) and F(p) with the corresponding
parts of the hard thermal loop (HTL) of the gluon self energy, the
latter being given by \cite{fren&tay}
\beq \delta \Pi^{ab}_{\mu \nu} (p) = -8N \delta^{ab} (2 \pi)^{-3}
     (f_{\mu \nu} - 4\pi \delta_{0 \mu} \delta_{0 \nu})
     {{\pi^2 g^2 T^2} \over 12}
\eeq
where
\beq f_{\mu\nu} (p)= 4\pi \int {{d\Omega} \over {4\pi}}
                     {{p_0 \hq_\mu \hq_\nu} \over {p\cd \hq}} \; ,
\eeq
and $\hq = (1, \vhq)$.
{}From this, with the Debye mass $m^2= g^2 T^2 N/3$, we have for the
effective $^*\!G$ and $^*\!F$ quantities
\beq \delta \Pi_{00} =- {{\vo p}^2 \over p^2}\, ^*\!F(p) =m^2 (1 - I(p)) \; ,
\eeq
with
\beq I(p)=\int{{d\Omega} \over {4 \pi}} {p_0 \over {p\cd\hq}} \; ,
\label{funcI}
\eeq
and
\beqa ^*\!G(p) &=& {1 \over 2} \left (\delta \Pi^\mu_\mu(p) +
                  {p^2 \over {\vo p}^2} \delta \Pi_{00}(p) \right )\nonumber \\
               &=& {m^2 \over 2} \left ({{p_0}^2 \over {\vo p}^2}
                  -{p^2 \over {\vo p}^2} I(p) \right ) \; .
\eeqa
Then $^*\!D_{ij}$ is
\beq ^*\!D_{ij}(p) = -{1 \over {p^2 - ^*\!G(p)}}
                  \left ( \delta_{ij}-{{p_i p_j} \over {\vo p}^2} \right )
                   -{1 \over {{\vo p}^2 +m^2 (1-I(p))}}
                   {{p_i p_j} \over {p_0}^2} \; .
\label{resumprop}
\eeq
To calculate the gluon self-energy, with two Feynman diagrams, we need
both the 3-gluon and the 4-gluon effective vertices. From
\cite{bra&pis,fren&tay},
\beq ^*\!\Gamma^{abc}_{0ij}(q,-p-q,p) = 2gif^{abc}
     \left (p_0 g_{ij} - m^2 \int {{d\Omega} \over {4\pi}}
     {{p_0 \hq_i \hq_j}\over{p\cd \hq (p+q)\cd \hq}}
     \right ) \; .
\label{resum3vert}
\eeq
Here we have set the external energy, $q_0$, to zero for the static
self-energy. Because of \eref{axiprop}, this is the only required 3-gluon
effective vertex component. For the 4-gluon effective vertex,
the six permutations of the gluon legs of the 4-gluon hard thermal
loop can be reduced to just a sum of two terms using identities
given in \cite{fren&tay}. For static self-energy and for
momentum configuration of the tadpole, these two terms give identical
contributions so for our purpose, the effective 4-gluon vertex is
\beq ^*\!\Gamma^{ab}_{00ij}(p,-p,q,-q) = (ig)^2 f^{adc} f^{bcd}
     \left (-2 g_{ij} + 4 m^2 \int {{d\Omega} \over {4\pi}}
     {{p_0 \hq_i \hq_j}\over{(p\cd \hq)^2 (p+q)\cd \hq}} \right )
     \; .
\label{resum4vert}
\eeq

\section{Leading Momentum Dependence of $\Pi_{00}(0,\vo q)$}

\label{leadingmom}

The leading term in the static self-energy is well known and is the
Debye mass squared. To find the leading correction and the momentum
dependent part, it is sufficient to consider only the zero mode in the
internal loop energy \cite{reb1} of the self energy. This is easy
to see by power counting of soft momenta. Take for example the
gluon self-energy graph with two 3-gluon vertices. The usual thermal
propagator in Euclidean space
\beq 1 \over {(2\pi nT)^2 + {\vo p}^2}
\eeq
means that for soft momentum and in the zero energy mode it goes as
$1/(gT)^2$ while for non-zero modes it is $1/T^2$. For a soft loop
self-energy graph, we then have $1/(gT)^4$ for two zero mode
propagators, $(gT)^2$ for momenta from the two vertices, $(gT)^4$ from
the loop 4-momentum, $g^2$ from the vertices and $1/g$ from the
distribution function. This yields $g(gT)^2$, which is $g$ times the
leading term. Similar soft power counting can be done to establish that
there is a contribution of the same order from the tadpole graph. For hard
loop, there appears to be $g(gT)^2$ terms left over from the leading
hard thermal loop $(gT)^2$ terms but they are proportional to
$\vo q\cd \vhq/\vmo Q$, where $\vo Q$ is the hard internal loop
momentum vector, and so vanish by virtue of the angular
integration.

With the effective quantities of the previous section, the
leading correction is
\beq \triangle \Pi_{00}^{ab}(0,\vo q)
     = \left \{ \triangle \Pi^{3g\;ab}_{00}(0, \vo q)
               +\triangle \Pi^{4g\;ab}_{00}(0, \vo q)
       \right \} \Bigr |_{zero \; mode} \; ,
\eeq
with
\beq \triangle \Pi^{3g\;ab}_{00}(0, \vo q) = {1 \over 2}
     T\sum_{p_0} \int_{soft} {{d^3p} \over {(2\pi)^3}} \,
     ^*\!\Gamma^{acd}_{0ij}(q,-p-q,p) ^*\!D_{jk}(p)
     ^*\!\Gamma^{bdc}_{0kl}(-q,-p,p+q) ^*\!D_{li}(p+q) \; ,
\label{cor3g}
\eeq
\beq \triangle \Pi^{4g\;ab}_{00}(0, \vo q) = {1 \over 2}
     T\sum_{p_0} \int_{soft} {{d^3p} \over {(2\pi)^3}} \,
     ^*\!\Gamma^{ab}_{00ij}(p,-p,q,-q) ^*\!D_{ij}(p)  \; .
\label{cor4g}
\eeq

As mentioned in the end of section \ref{linresp}, we need only the
small $\vo q$ limit of the static self-energy when looking for a
possible odd term in $\vmo q$ \cite{baier&kal,gale&kap}.
This simplifies our problem considerably. In fact we can drop the bare
vertex part of \eref{resum4vert} because its contribution is independent
of $\vo q$.

We note that since, on the one hand, the effective vertex
\eref{resum3vert} and the hard thermal loop part of \eref{resum4vert}
are proportional to $p_0$ and, on the other hand, the transverse part
of the gluon propagator has no singular factor in $p_0$,
the zero mode part of the transverse-transverse gluon
contribution of \eref{cor3g} vanishes. The same is true for the
transverse gluon contribution of \eref{cor4g}.

For the transverse-longitudinal gluon contribution, it turns out
that, in the zero mode, the bare-HTL vertex and the HTL-HTL vertex
combinations are also zero because
\beq \int {{d\Omega} \over {4\pi}}
     \left ( \delta_{ij}-{{p_i p_j} \over {\vo p}^2} \right )
     {{(p+q)_i \hq_j} \over {\vo p \cd \vhq}} =0 \; ,
\eeq
and
\beq \int {{d\Omega_1} \over {4\pi}} {{d\Omega_2} \over {4\pi}} \left (
     {{\vhq_1 \cd \vhq_2} \over
      {\vo p \cd \vhq_1 \vo p \cd \vhq_2}}
     -{1 \over {\vo p}^2} \right ) = 0 \; ,
\eeq
where we integrate over the directions of $\vhq_i$ through
$d\Omega_i$, $i=1,2$. The remaining bare-bare vertex
combination does not give any odd power of $\vmo q$: the integral
in $\vo p$ reads
\beq \int {{d^3 p} \over {(2\pi)}^3}
     \left ({\vo p \cd \vt pq}^2 - {\vo p}^2 {\vt pq}^2 \right )
     {1 \over {({\vo p}^2)^2 ({\vt pq}^2 +m^2)} } \; .
\label{barbartranlong}
\eeq
Because the denominator of \eref{barbartranlong} is protected by the
mass from infra-red divergence, we can expand at small
$\vmo q < m$. It is clear that odd powers of $\vmo q$ always come with
odd powers of $\cos \theta$, where $\theta$ is the angle between $\vo q$
and $\vo p$, therefore no odd power of $\vmo q$ appears in the small
$\vmo q$ limit of the longitudinal-transverse contribution.

Before discussing the longitudinal-longitudinal contribution, it must
be mentioned that the Matsubara frequency sum is now ill defined
for certain terms because of the presence of two longitudinal
propagators: each has a $1/p^2_0$ factor and we have only a
factor of $p_0$ from each vertex. Therefore terms which do not have any
extra factor of $p_0^i,\;i\ge 2$ in the numerator are singular.

In these cases, as was done in \cite{kaj&kap,hei&kaj&toi}, we will
perform the frequency sums by contour integrals using the usual equation
\cite{kap}:
\beq T\sum_n f(p_0)={1\over {2i\pi}} \Big \{
     \int^{i\infty}_{-i\infty} dz f(z) +
     \int^{i\infty+\epsilon}_{-i\infty+\epsilon} dz
     \Big (f(z)+f(-z) \Big ) {1 \over {e^{z/T}-1}} \Big \}
\label{ansatz1}
\eeq
This equation is valid only if $f(p_0)$ has no singularities
on the imaginary axis. This is not the case if $f(p_0)$ contains
$1/p^2_0$ or $1/p_0$ factors. We therefore employ the usual ansatz,
namely, that the sum is still given by the above integral. Thus the
$1/p_0$ and $1/p^2_0$ poles, if existing on their own with no other poles,
are vanishing.
\beq T \sum 1/p_0 = 0 \mbox{\rm \hskip1cm , \hskip 1cm}
     T \sum 1/p^2_0 =0 ; .
\label{ansatz2}
\eeq

Let us point out that if one takes the finite T axial gauge propagator
directly from zero T (i.e. that is one ignores the problem
with periodicity of the longitudinal propagator at finite T
\cite{james&land}, see comment at the end of section \ref{conclude}) then
this is the only known way at finite T to treat the
$1/p^2_0$ singularities. The alternative suggested by Leibbrandt
and Staley \cite{leib&stal} amounts to performing the uniform replacement
\beqa {1 \over p_0} & \rightarrow & \lim_{\epsilon \rightarrow 0}
      {p_0 \over {p_0^2 + i\epsilon}} \; , \; \epsilon > 0
      \nonumber \\
      {1 \over p^2_0} & \rightarrow & \lim_{\epsilon \rightarrow 0}
      {p^2_0 \over {(p_0^2 + i\epsilon)}^2} \; , \; \epsilon > 0
\eeqa
and dropping the zero mode.

The only justification of this prescription is that it leads to the
correct answer for the leading order Debye mass. This is however
not surprising since this calculation is only sensitive to hard loop
momenta. Furthermore, it is known that due to the plasmon
effect \cite{kap}, the leading correction to $m^2$ starts at
$O(g^3)$ and not at $O(g^4)$ as is expected in
naive perturbation theory. We know by using
soft power counting that the $O(g^3)$ correction comes only
from the zero mode, therefore we argue that any prescription
which removes the zero mode cannot be the correct one.

We calculate the longitudinal-longitudinal contribution by
expanding the longitudinal propagator, making
use of $m^2/(\vo p^2 + m^2) < 1$. We obtain from \eref{resumprop}
the expanded form
\beq ^*\!D_{L\,ij}(p) = - {{p_i p_j} \over {\vo p^2 +m^2}}
     \left ({1 \over {p_0^2}} \right )  \sum_{n =0}^\infty
     \left ({{m^2 I(p)} \over {\vo p^2 +m^2}} \right )^n \; .
\label{longpropexp}
\eeq
The longitudinal-longitudinal contribution is then
\beqa \Delta \Pi_{00\;LL}^{3g\;ab}(0, \vo q)= \!\!\!\!&&\!\!\!\!
      2g^2 NT\delta_{ab} \int {{d^3 \vo p} \over {(2\pi)^3}}
      {1 \over {({\vo p}^2+m^2) ((\vo p+\vo q)^2+m^2)}} \nonumber \\
      \!\!\!\!&&\!\!\!\!
      \sum_{n_1=0}^\infty \sum_{n_2=0}^\infty \sum_{p_0} {1 \over p_0^2}
      \left ({{m^2 I(p)} \over {\vo p^2 +m^2}} \right )^{n_1}
      \left ({{m^2 I(p+q)} \over {(\vo p+\vo q)^2 +m^2}} \right )^{n_2}
      \nonumber \\ \!\!\!\!&&\!\!\!\!
      \Big \{ \vo p\cd (\vo p+\vo q)+m^2-m^2(I(p)+I(p+q))
             +m^2 p_0^2 J(p,p+q) \Big \}^2
\label{fulonglong}
\eeqa
where $I(p)$ is given in \eref{funcI} and $J(p,q)$ is
\beq J(p,q) = \int {{d\Omega} \over {4\pi}} {1 \over {p\cd \hq \, q\cd \hq}}
     \; .
\eeq
(for more details of this function, see \cite{fren&tay}). As the function
$I(p)$ is proportional to $p_0$, all frequency sums in \eref{fulonglong}
with $n_1+n_2> 2$ have vanishing zero mode contributions.
We are left with the following terms in \eref{fulonglong}:
\begin{itemize}
\item one frequency sum with a singular zero mode proportional to
      $1/p_0^2$, which vanishes according to the ansatz \eref{ansatz2}.
\item frequency sums with zero mode proportional to $1/p_0$. These
      are performed using \eref{ansatz1} and \eref{ansatz2}. The
      Bose-Einstein distribution function is approximated
      $N(p) \simeq T/p$\ for $p<<T$, in order
      to extract the leading contributions. These, for the same reason as
      \eref{barbartranlong}, yield no odd power of $\vo q$ in the
      $\vo p$-integrals.
\item frequency sums with well-defined zero modes: we perform the
      angular integrals first and then take the zero mode. The only
      odd terms in $|\vo q|$ originate from the product $I(p)I(p+q)$
      proportional to
\beq \int {{d\Omega_1} \over {4\pi}}{{d\Omega_2} \over {4\pi}}
     {1 \over {p\cd \hq_1}} {1 \over {(p+q)\cd \hq_2}} =
     {{\ln \bigg ({{p_0+\vmo p}\over{p_0-\vmo p}} \bigg )
       \ln \bigg ({{p_0+\vmt pq}\over{p_0-\vmt pq}} \bigg )}
       \over {4\vmo p \vmt pq}}
     {\longrightarrow \atop _{p_0=0}}
     - {\pi^2 \over 4} {1 \over {\vmo p \vmt pq}} \; .
\label{crufac}
\eeq
\end{itemize}
Collecting the latters and adding an analogous contribution coming
from the longitudinal part of the tadpole diagram \eref{cor4g}, we get
\beqa -{{g^2 NT m^4 \pi^2} \over 2}\!\int\! {{d^3 p} \over {(2\pi)}^3}
      \!\!&&\!\!
     {1 \over {{\vmo p}{\vmt pq}({\vt pq}^2 +m^2)^2 ({\vo p}^2 +m^2)^2}}
     \nonumber \\ \!\!&&\!\!
     \times \left ( {{(\vo q^2)}^2 \over 2}+{\vo q}^2 ({\vo p}^2+m^2)
                   -{(\vo q\cd\vo p)}^2 \right ) \; .
\label{crufacterm}
\eeqa
{}From this we find no linear term but a cubic term in $\vmo q$
which is $g^2NT {\vmo q}^3/8m^2$. So that, in the small $\vmo q$ limit,
we get the following behaviour for $\Pi_{00}$:
\beq \lim_{\vmo q \rightarrow 0} \Pi_{00}(0, q) = m^2
     +\alpha g^2 T \vmo q + \beta {{g^2 T N {\vo q}^2} \over m}
     +\gamma {{g^2 TN {\vmo q}^3} \over m^2} + O({\vmo q}^4)
\label{expselfeng}
\eeq
where $\alpha$, $\beta$ and $\gamma$ are numerical factors with
$\alpha =0$ ,$\gamma = 1/8$. The factor $\beta$ is not as
important in front of the $\vo q^2$ term already present in the
potential \eref{potent} since it is smaller by a factor of $g$.
Had the factor \eref{crufac} not been present, no odd $\vmo q$
term would have appeared. The positive cubic term in \eref{expselfeng}
is responsible for the algebraic decrease of the potential $V(r)$
of \eref{Eucpotent}:
\beq V(r\rightarrow \infty) \simeq
     {{N^2-1} \over {4\pi^2}} {{3 g^4(T) T} \over {(r m)}^6}
\eeq
which behaves as $\sim 1/r^6$ in agreement with \cite{baier&kal}.
The presence of this cubic term means that the screening potential
is no longer dominated by a pole but by some cut starting at $\vmo q=0$.
On a physical level, the self-energy in TAG is very different from that
in covariant gauge: we get a longitudinal-longitudinal contribution
whereas in covariant gauge, this contribution is zero.
We find that we have no need to include the magnetic mass since
transverse gluon does not yield any odd term in the zero mode.

\section{Discussion and Conclusion}

\label{conclude}

In this paper, using Braaten and Pisarski resummed perturbation
theory we study the leading correction to the Debye mass in TAG.
Indeed, it is the only gauge where the potential between two charges
is unambiguously related to the two-point function.
We find that provided there is no branch cut touching the
real axis in the first quadrant of the complex q plane, one only
requires the time-time component of the gluon self-energy at
small $\vo q$. This gives a cubic term which yields a
repulsive power screening potential for the non-abelian
theory. This is in direct contrast with the accepted notion of
Debye mass. Nevertheless, one has
to be careful in trusting such a result. We have merely looked at
the small $\vmo q$ behaviour of $\Pi_{00}(0,\vo q)$ for the
dominant behaviour of the potential. We expect the
same pole as found by Rebhan \cite{reb1,reb2} to be present in
this gauge due to the theorem \cite{kob&kun&reb} that the poles of
the propagator are gauge invariant. But in this case, the behaviour of
the potential is not dominated by this pole and therefore it could not
be interpreted as the Debye mass.

Attempts to understand the difference of our result with other gauges
may lead one to question the use of the ansatz summarised in
\eref{ansatz2} which is known to be true up to subleading order in the
non-resummed theory \cite{hei&kaj&toi} and is untested at higher order.
However, notice that the crucial cubic term in $\Pi_{00}$ comes from
frequency sum with well-defined zero mode terms.

Also, James and Landshoff \cite{james&land} have pointed out that the
TAG imaginary time longitudinal propagator should not be
periodic in time as in the usual Matsubara framework used in this paper:
the requirement to include only physical states in the thermal
a\-verage breaks periodicity. They have initiated a rather complex
formalism appropriate to this gauge based on this. Their new
longitudinal propagator has the advantage that it is free of
$p_0=0$ pole and therefore one does not have to use any pragmatic
ansatz. Calculating the correction to the Debye mass in this
formalism is necessary to confirm our result.

After finishing this work, we noticed Braaten and Nieto
\cite{bra&nie} and Rebhan \cite{reb4} have extracted the Debye
mass from the gauge invariant correlator of two Polyakov loops.
They found the expected exponential behaviour of the potential
and no power screening. We do not know exactly where is the source of
this disagreement. It could be due to one of the problems that we have
already discussed above. We leave this as further investigation.

\vskip 0.8cm
\noindent{\bf Acknowledgements}
\vskip 0.5cm

S.P. would like to thank J.P. Leroy and E. Pilon for discussions.
We would like to thank R. Baier, P.V. Landshoff, H. Nachbagauer,
A.K. Rebhan and D. Schiff for discussions. We would also like to
thank D. Schiff for a critical reading of the
manuscript. S.M.H.W acknowledges financial support from the
Leverhulme Trust.

\end{document}